\documentclass{egpubl}
\usepackage{eg2026}
\makeatletter
\let\ps@titlepage\ps@empty   
\makeatother

\usepackage[T1]{fontenc}
\usepackage{dfadobe}  

\usepackage{cite}  
\BibtexOrBiblatex
\electronicVersion
\PrintedOrElectronic
\ifpdf \usepackage[pdftex]{graphicx} \pdfcompresslevel=9
\else \usepackage[dvips]{graphicx} \fi

\usepackage{egweblnk} 
\usepackage{amsmath}
\usepackage{amsfonts}
\usepackage{algorithm}
\usepackage{algorithmic}
\usepackage{graphicx}
\usepackage{subcaption}
\usepackage{booktabs}
\usepackage[dvipsnames]{xcolor}

\title[EG \LaTeX\ Author Guidelines]%
      {Interpolated Adaptive Linear Reduced Order Modeling for Deformation Dynamics}

\author[Y. Tao, M. Chiaramonte, P. Fernandez]%
{\parbox{\textwidth}{\centering
Yutian Tao$^{1,2}$,\;
Maurizio Chiaramonte$^{2}$,\;
Pablo Fernandez$^{2}$\\[4pt]
$^{1}$University of Wisconsin–Madison, USA\\
$^{2}$Meta, USA\\
}}

\begin{document}

\teaser{
 \includegraphics[width=1.0\linewidth]{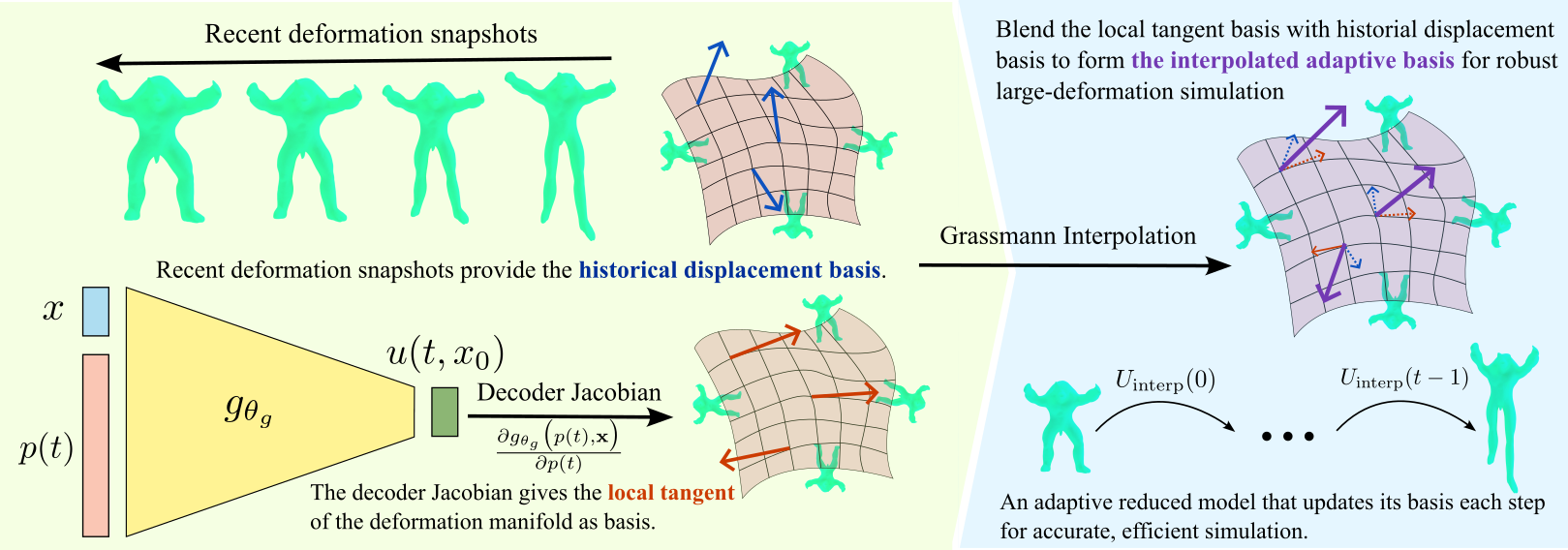}
 \centering
 \caption{We propose a method that forms an adaptive basis for robust, efficient simulation under large deformations by blending the local tangent of the deformation manifold—learned from the decoder’s gradient—with recent displacement history. Our approach achieves 40–70\% lower online error than PCA-based ROM, with per-step costs only 1.3–1.5× higher, while maintaining interactive runtimes.}
\label{fig:teaser}
}

\maketitle
\begin{abstract}
    Linear reduced-order modeling (ROM) is widely used for efficient simulation of deformation dynamics, but its accuracy is often limited by the fixed linearization of the reduced mapping. We propose a new adaptive strategy for linear ROM that allows the reduced mapping to vary dynamically in response to the evolving deformation state, significantly improving accuracy over traditional linear approaches. To further handle large deformations, we introduce a historical displacement basis combined with Grassmann interpolation, enabling the system to recover robustly even in challenging scenarios. We evaluate our method through quantitative online-error analysis and qualitative comparisons with principal component analysis (PCA)-based linear ROM simulations, demonstrating substantial accuracy gains while preserving comparable computational costs.


\end{abstract}  

\section{Introduction}
Efficient and accurate simulation of deformable objects remains a central challenge in computer graphics, with applications in animation, virtual and augmented reality, and robotics.  
A common strategy to achieve interactive performance is reduced-order modeling (ROM), which projects high-dimensional dynamics onto a low-dimensional subspace, dramatically reducing the number of degrees of freedom (DOFs) while preserving important features.  
Traditional linear subspace models—constructed from modal analysis or data-driven bases such as PCA—are especially attractive for their simplicity and speed.  
However, their accuracy is fundamentally constrained by linearization: when deformations are large or highly nonlinear, a fixed reduced mapping cannot capture the evolving state, leading to drift, loss of detail, and inability to recover to the undeformed rest configuration.

Researchers have explored nonlinear and data-driven approaches that learn more expressive low-dimensional manifolds to overcome these limitations.  
Autoencoder-based methods and implicit neural representations demonstrate that compact latent spaces can capture complex deformations beyond the reach of linear subspaces.  
Among these, Continuous ROM (CROM) \cite{chen2022crom} leverages encoder–decoder neural fields to map between latent coordinates and continuous displacement fields, offering discretization flexibility and improved fidelity.  
Despite their expressiveness, many nonlinear ROMs incur nontrivial per-step costs—such as evaluating full-order residuals or Jacobians—which limit their practical speed.

A complementary direction introduces adaptivity into the reduced space itself.  
Basis enrichment and localized corrections can recover accuracy when previously unseen deformation patterns arise, balancing efficiency and fidelity.  
However, triggering and integrating these adaptations robustly remains challenging; naive heuristics may over-refine, introduce instabilities, or erode the computational advantage.

To address these challenges, we propose an interpolated adaptive linear ROM for deformation dynamics that combines the efficiency of linear subspaces with a decoder-based representation inspired by CROM, enabling a time-varying reduced mapping while remaining compatible with continuous neural frameworks.  
Our method derives a time-dependent reduced basis from the decoder Jacobian (gradient) evaluated at the current latent state, which naturally aligns the subspace with the local tangent of the learned nonlinear manifold while retaining the computational structure of linear ROM.  
To further improve robustness under large deformations, we introduce a historical displacement basis and blend it with the current adaptive basis through Grassmann interpolation, which smoothly transitions between the two sets of basis vectors—like morphing one subspace into another—while preserving their geometric structure and capturing important features from both recent history and the current deformation.  
This history-aware interpolation enables accurate recovery to the rest state after extreme deformations, overcoming the drift that arises when using only the decoder gradient. In summary, our main contributions are:
\begin{itemize}
  \item \textbf{Adaptive linear subspace method:} Extends CROM by efficiently using the decoder Jacobian to construct a time-varying reduced mapping aligned with the evolving deformation state, adding only negligible overhead to the standard ROM pipeline.
  \item \textbf{History-aware Grassmann interpolation:} Blends a historical displacement basis with the current adaptive basis to enable perfect recovery to the undeformed rest configuration even under large deformations.
  \item \textbf{Comprehensive evaluation:} Demonstrates \textbf{40--70\%} lower online error than PCA-based ROM while maintaining interactive runtime with per-step costs within \textbf{1.3--1.5$\times$} that of PCA, supported by quantitative metrics and qualitative visual comparisons across diverse objects and challenging scenarios.
\end{itemize}
\section{Related Work}
\noindent\textbf{Linear Subspace Model Reduction.}
Linear subspace model reduction has long been a standard approach for accelerating physical simulations in computer graphics and engineering, particularly in deformation dynamics. Early work on modal analysis~\cite{pentland1989good} showed that an object's motion could be efficiently represented as a superposition of a small number of vibration modes, later extended with constraints~\cite{hauser2003interactive}, modal warping for large rotations~\cite{choi2005modal}, and modal derivatives that enrich the basis with quadratic mode couplings to capture large deformations~\cite{barbivc2005real}. These extensions established modal analysis as a practical real-time tool for deformation dynamics. At the same time, data-driven \emph{Principal Component Analysis} (PCA)--based subspace methods emerged. Rather than relying on vibration modes alone, PCA is applied to simulation snapshots to extract dominant deformation patterns. Early papers demonstrated the feasibility of data-driven reductions for elastostatics~\cite{james1999artdefo}, character animation~\cite{kry2002eigenskin}, and finite element simulations of soft bodies~\cite{krysl2001dimensional}, where PCA or other snapshot-based decompositions were used to construct low-dimensional bases.

Beyond deformation, linear subspace reduction has been successfully applied to other domains in graphics. In fluids, PCA-derived flow bases enable real-time incompressible simulation~\cite{treuille2006model}. Linear subspaces also support reduced sound simulation~\cite{o2002synthesizing,james2006precomputed}, shape deformation~\cite{von2013efficient,brandt2016geometric}, and computational design~\cite{xu2015interactive,ulu2017lightweight,musialski2016non}. However, these methods suffer from critical accuracy limitations: if a deformation or motion lies outside the span of the chosen subspace basis, the reduced simulation simply cannot represent it.

\noindent\textbf{Model Reduction with Neural Networks.}
Recent years have seen rapid progress in machine learning approaches for reduced-order modeling. These methods use neural networks to construct compact nonlinear manifolds that serve as effective reduced simulation spaces. A main class of methods relies on autoencoders to encode deformation snapshots into low-dimensional latent vectors. Nonlinear manifold reduction for PDEs was introduced through autoencoders~\cite{lee2020model}, and this concept was extended to deformable solids by learning latent dynamics in reduced coordinates~\cite{fulton2019latent}. To enhance accuracy for highly nonlinear deformation, a deep autoencoder with high-order differentiability---enabling computation of Hessian-based dynamics---was proposed~\cite{shen2021high}. Correction techniques further refine these models; for instance, target-specific adjustments have been proposed to handle contact phenomena~\cite{romero2023learning}.  
Beyond ROM for solids, neural networks have been applied to a wide range of graphics problems to learn expressive low-dimensional latent spaces. Latent-space physics frameworks use autoencoders and recurrent networks to evolve fluid states entirely in a low-dimensional space, achieving dramatic speedups while retaining physical accuracy~\cite{kim2019deep,wiewel2019latent}. Similar principles have been applied to deformable bodies, where neural subspace dynamics enable fast interactive simulation by learning physical updates directly in a reduced space~\cite{holden2019subspace,li2025self}. In character animation, periodic autoencoders learn compact motion-phase manifolds for realistic motion synthesis~\cite{starke2022deepphase}.

A complementary path uses implicit neural representations, enabling discretization-independent model reduction. Some frameworks explore mesh-agnostic mappings that predict piecewise linear deformations without requiring shared triangulations~\cite{aigerman2022neural}. Continuous Reduced-Order Modeling proposed a continuous neural-field formulation for PDEs~\cite{chen2022crom}, and subsequent work explored discretization-free subspace representations that allow latent dynamics to be learned independently of the underlying mesh~\cite{chen2023model}, linear continuous reduction models that directly learn spatially smooth mappings from undeformed to deformed states~\cite{chang2023licrom}, and neural stress field reduction that encodes high-dimensional stress responses into compact implicit neural fields~\cite{zong2023neural}. Together, these approaches highlight the versatility of neural latent representations in graphics, which can operate not only on displacements but also on internal quantities such as strain, stress, fluid velocity, or motion phases. While these neural field--based methods generally achieve higher accuracy and flexibility than linear subspace models, they also tend to be more computationally demanding than traditional linear reduction methods.

\noindent\textbf{Adaptive Subspace Methods for Dynamics.}
A major limitation of fixed subspace simulation is the inability to represent unanticipated local deformations or contact events. To address this, a variety of adaptive subspace methods have been proposed that dynamically update the reduced basis during simulation time. One family of approaches focuses on global basis expansion. An early example is the online reduction strategy, which interleaves subspace solves with occasional full-space solves to enrich the basis whenever the reduced model diverges significantly from the ground truth~\cite{kim2009skipping}. This strategy minimizes expensive full solves while adaptively expanding the snapshot basis. In cloth simulation, a related technique precomputes a large dictionary of basis vectors and adaptively selects subsets depending on the garment’s pose, thereby reproducing diverse folding patterns with a small active subspace~\cite{hahn2014subspace}.

Another direction is localized subspace enrichment, where the reduced basis is extended only in regions of high deformation or contact.  
A notable method augments modal bases with localized displacement fields generated on demand at collision sites~\cite{harmon2013subspace}.  
Hybrid approaches further combine subspace simulation with learned corrections: neural models can predict contact-driven deformations and inject them into a handle-based subspace, efficiently capturing both broad dynamics and fine-scale contact detail~\cite{romero2021learning,romero2022contact}.  
Adaptive rigidification offers a different perspective by dynamically reducing the number of elastic degrees of freedom and treating nearly static regions as rigid to accelerate computation~\cite{mercier2022adaptive}.  
The Trading Spaces method generalizes these families by introducing an oracle-driven time integration scheme that dynamically switches between subspace simulation, nodal enrichments, and even full-space solves as necessary~\cite{trusty2024trading}.

These adaptive methods generally deliver substantial speedups compared to fixed reduced spaces while improving accuracy for large deformations and contact.  
However, they also face challenges.  
Oracles used to decide when and where to adapt are often simplistic, focusing only on energy error~\cite{kim2009skipping} or applying enrichment broadly at detected contacts~\cite{harmon2013subspace,teng2015subspace}, which can either miss important events or lead to over-refinement.  
Furthermore, switching between models on a per-time-step basis can produce popping artifacts if transitions are not handled consistently.  
Unlike these enrichment strategies, our method adapts the reduced mapping continuously by following the local gradient of a learned decoder, avoiding the need for explicit oracle-based enrichment.

\section{Technical Background}
In this section we review the two principal ROM approaches that underpin our work: linear ROM and CROM.  
Both have been widely used for simulating soft-body dynamics, yet they differ fundamentally in their assumptions, expressiveness, and computational trade-offs.

\begin{figure}[htpb]
    \centering
    \includegraphics[width=0.45\textwidth]{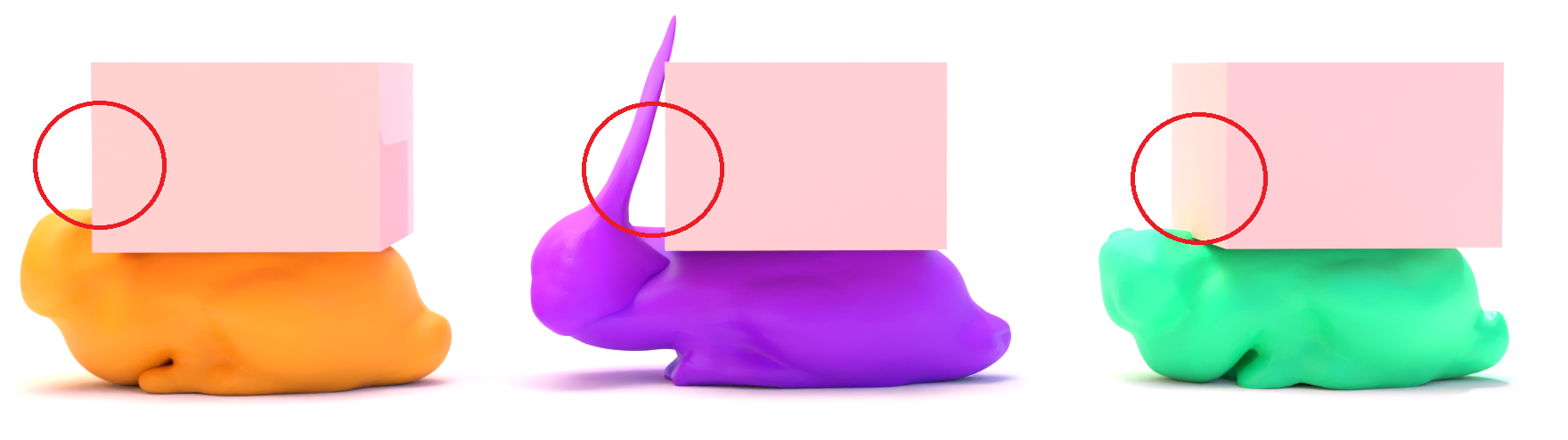}
    \caption{
    Bunny compression under large deformation: \color{violet}PCA\color{black}\ (middle) exhibits noticeable artifacts compared to \color{orange}FOM\color{black}\ and \color{ForestGreen}ours\color{black}.
    }
    \label{fig:bunny_comp_img}
\end{figure}

\subsection{Linear Reduced-Order Modeling}

The simulation of deformable soft bodies is a core problem in computer graphics and computational mechanics.  
When discretized using the finite element method (FEM), the motion of a soft body is governed by Newton’s second law
\begin{equation}
    M \ddot{u}(t) = f_\mathrm{int}\big(u(t)\big) + f_\mathrm{ext}(t),
    \label{eq:fom}
\end{equation}
where $u(t) \in \mathbb{R}^{3n}$ is the vector of nodal displacements for a mesh with $n$ vertices,  
$M \in \mathbb{R}^{3n \times 3n}$ is the mass matrix, $f_\mathrm{int}(u(t))$ is the generally nonlinear internal force,  
and $f_\mathrm{ext}(t)$ represents external forces such as gravity, collisions, or user interactions.  
For realistic models $n$ can reach tens of thousands, making full-order model prohibitively expensive for interactive applications.

Linear ROM alleviates this cost by approximating the high-dimensional state $u(t)$ with a much lower-dimensional representation.  
The core idea is to project the dynamics onto a fixed linear subspace spanned by a small set of basis vectors,
\begin{equation}
    u(t) = U q(t),
    \label{eq:linear}
\end{equation}
where $U \in \mathbb{R}^{3n \times r}$ contains the basis vectors as columns and $q(t) \in \mathbb{R}^{r}$ are the reduced coordinates with $r \ll n$.  

The basis $U$ is typically constructed in a data-driven manner.  
Principal Component Analysis (PCA) is commonly applied to a set of simulated or captured deformations to extract the dominant modes of variation.  
These modes capture the most representative deformation patterns and provide an efficient subspace for runtime simulation.

Substituting~\eqref{eq:linear} into~\eqref{eq:fom} and premultiplying by $U^{T}$ yields the reduced equations of motion
\begin{equation}
    \tilde{M} \ddot{q}(t) = \tilde{f}_\mathrm{int}\big(q(t)\big) + U^{T} f_\mathrm{ext}(t),
\end{equation}
where $\tilde{M} = U^{T} M U$ is the reduced mass matrix and  
$\tilde{f}_\mathrm{int}(q) = U^{T} f_\mathrm{int}(U q)$ is the reduced internal force.

This reduction dramatically lowers computational cost and enables real-time simulation and control of soft bodies.  
Linear ROM performs well when deformations remain moderate and can be represented accurately within a fixed subspace, such as small vibrations, bending, or stretching.  
However, accuracy degrades when the system exhibits strongly nonlinear behavior or large deformations as shown in Figure \ref{fig:bunny_comp_img}, since the fixed basis cannot capture the evolving dynamics.  
Moreover, because the basis is tied to a particular mesh, changes in topology or even vertex sampling require regenerating the basis, limiting flexibility and reusability.  
For a broader review of linear ROM in graphics and mechanics, see~\cite{sifakis2012fem}.

\subsection{Continuous Reduced-Order Modeling}

To overcome the limitations of fixed linear subspaces, recent work has introduced nonlinear and data-driven approaches.  
Continuous Reduced-Order Modeling (CROM) represents the deformation field with an implicit neural function that maps low-dimensional latent coordinates directly to continuous displacements.

\begin{figure}[htbp]
  \centering
  \includegraphics[width=0.45\textwidth]{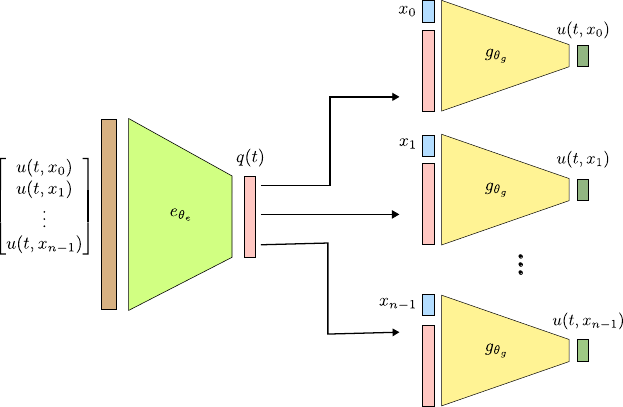}
  \caption{Continuous reduced-order modeling (CROM).  
  The encoder $e_{\theta_e}$ maps a full displacement field $u(t, \mathbf{x})$ to latent coordinates $q$, and the decoder $g_{\theta_g}$ reconstructs continuous displacements from $q$ and spatial position $x_i$.}
  \label{fig:crom}
\end{figure}

In CROM, the displacement at a material point $x_i \in \mathbb{R}^3$ is expressed as
\begin{equation}
    u(t,x_i) = g_{\theta_g}\big(q(t), x_i\big),
\end{equation}
where $q(t) \in \mathbb{R}^r$ is a latent vector and $g_{\theta_g}$ is a neural decoder with parameters $\theta_g$.  
An encoder $e_{\theta_e}$ maps a high-dimensional displacement field to the latent coordinates, while the decoder reconstructs displacements from $q$ and $x$, as illustrated in Figure~\ref{fig:crom}.  
The networks are trained jointly to minimize reconstruction error over a dataset of simulated deformations, after which the reduced dynamics can be integrated entirely in the latent space.

By operating on continuous spatial coordinates rather than a fixed discretization, CROM learns a latent manifold for the deformation field itself.  
This design allows training on heterogeneous meshes or sampling patterns and supports adaptive remeshing or variable-resolution evaluation at runtime.  
It also provides strong memory efficiency, since the latent dimension scales with the intrinsic complexity of the deformation rather than the number of mesh nodes, and experiments have shown order-of-magnitude speedups and accuracy improvements over PCA and autoencoder baselines.

Despite these advantages, CROM is not without limitations.  
Its expressiveness is bounded by the manifold captured during training and it cannot guarantee accurate prediction for dynamics outside that range.  
Each time step still requires evaluating the neural decoder and sampling the original PDE at a reduced set of spatial points, so CROM does not completely eliminate expensive PDE solves.  
In addition, obtaining a high-quality latent model demands an offline training phase on full-order simulation data, which can be computationally expensive.

Like other learning-based ROM, our approach requires high-quality training data and remains limited by the coverage of the learned deformation manifold,  
but within that manifold it offers a far better trade-off between efficiency and expressiveness than classical PCA-based ROM.  
While the adaptive basis is obtained from the gradient of the CROM decoder, which may differ from the theoretically optimal subspace,  
this computation is extremely fast and adds negligible overhead compared with the cost of assembling or evaluating a full reduced system.  
By combining this fast gradient-based adaptation with a time-updated subspace, our method delivers rich nonlinear accuracy and real-time performance in a simple, linear ROM framework.

\section{Proposed Method}
In this section, we presents our method for deformation dynamics. We first describe the adaptive model-reduction framework, then detail the latent-space dynamics that govern reduced simulations, and finally introduce a history-aware interpolation strategy for handling large deformations.

\subsection{Adaptive Linear Model Reduction}

Before introducing adaptivity, we first describe how our implementation modifies the traditional linear reduced-order model (ROM).  
Instead of directly applying Equation~\ref{eq:linear}, we adopt an incremental formulation
\begin{equation}
    u(t) = u(t-1) + U q(t),
\end{equation}
where the displacement is updated relative to the previous step.  
This incremental form improves numerical stability and accuracy in scenarios with large deformations, where a direct global update may accumulate significant error.

To incorporate adaptivity, we generalize the basis matrix $U$ to be time dependent:
\begin{equation}
    u(t) = u(t-1) + U(t-1) q(t).
    \label{eq:adapt}
\end{equation}
Here $U(t-1)$ is updated at every step to better capture the local deformation behavior.  
Because CROM represents deformations on a nonlinear manifold while our incremental update is formulated in a linear space, each state is projected onto the CROM manifold using the encoder
\begin{equation}
    p(t) = e_{\theta_e}\big(u(t)\big),
\end{equation}
where $p(t) \in \mathbb{R}^r$ is the latent variable with the same dimension $r$ as $q(t)$.

The adaptive basis $U(t)$ is computed from the CROM decoder by differentiating its output with respect to the latent variable.  
Let $g_{\theta_g}\big(p(t),\mathbf{x}\big)\in\mathbb{R}^{3n}$ denote the stacked displacement field obtained by evaluating the decoder at all mesh nodes $\mathbf{x}=\{x_i\}_{i=1}^{n}$.  
Since $g_{\theta_g}(p(t),x_i)$ returns the displacement at a single spatial location $x_i$, the global basis matrix is assembled by concatenating the per-node decoder Jacobians:
\begin{equation}
    U(t)\;\approx\;
    \frac{\partial g_{\theta_g}\big(p(t),\mathbf{x}\big)}{\partial p(t)}
    \;\in\;\mathbb{R}^{3n\times r}.
\end{equation}
This Jacobian aligns the reduced basis with the local tangent of the learned deformation manifold, enabling the adaptive model to capture strongly nonlinear dynamics while retaining the computational structure of a linear subspace.

The procedure is applied at every time step as shown in Algorithm \ref{alg:adapt}: the current state is projected to the CROM latent space, the decoder gradient provides the updated basis, the reduced variables $q(t)$ are advanced by solving the latent dynamics, and the full displacement $u(t)$ is incrementally reconstructed.  
This adaptive construction produces a time-varying basis that tracks the evolving geometry of the deformation, enabling efficient yet accurate simulation of complex soft-body motions.

\begin{algorithm}[htpb]
\caption{Adaptive Linear Model Reduction Procedure}
\label{alg:adapt}
\begin{algorithmic}[1]
\STATE \textbf{Input:} Initial displacement $u(0)\in\mathbb{R}^{3n}$
\FOR{each time step $t = 1,2,\ldots$}
    \STATE $p(t-1) \leftarrow e_{\theta_e}\big(u(t-1)\big)$
    \STATE $U(t-1) \leftarrow 
           \displaystyle
           \frac{\partial g_{\theta_g}\big(p(t-1),\mathbf{x}\big)}{\partial p(t-1)}$
    \STATE Compute $q(t)$ using Newton solver (Algorithm~\ref{alg:newton})
    \STATE $u(t) \leftarrow u(t-1) + U(t-1)\,q(t)$
\ENDFOR
\end{algorithmic}
\end{algorithm}

\subsection{Latent Space Dynamics}
With the adaptive basis in place, the reduced coordinates $q(t)$ evolve directly in the latent space to capture deformation dynamics efficiently.  
At each time step, after projecting the current state onto the CROM manifold and updating $U(t)$, we solve for $q(t)$ by minimizing the total system energy—including elastic, inertial, gravitational, and contact terms—within the current adaptive subspace:
\begin{equation}
    q(t) = \arg\min_{q} E\big(u(q)\big),
\end{equation}
where $E$ is the total energy evaluated at the configuration $u(q)$.
In our method, we employ implicit time integration, formulating the energy $E$ to include elastic potential, inertia, contact, and friction terms as required by the application. Specifically, we adopt the implicit energy-diminishing integrator\cite{stuart1998dynamical}, and handle contact and friction using the incremental potential contact method \cite{li2020incremental}. The resulting minimization problem can be solved using standard numerical techniques, such as Newton's method or other gradient-based optimization algorithms, depending on the system's nonlinearity. In our implementation, we utilize a Newton iteration of the form:
\begin{equation}
    q^{k + 1} = q^{k} - \alpha^k \left[\nabla^2_q E\big(u(q^{k})\big)\right]^{-1} \nabla_q E\big(u(q^{k})\big),
\end{equation}
where $\alpha^k$ is a step size determined by a line search or set to $1$ for standard Newton's method. By applying the chain rule, the reduced gradient and Hessian can be expressed as
\begin{align}
    \nabla_{q} E\big(u(q^{k})\big)
    &= \left( \frac{\partial u}{\partial q} \right)^{\!T}
       \nabla_{u} E\big(u(q^{k})\big)
       = U^{T} \nabla_{u} E\big(u(q^{k})\big), \\
    \nabla^{2}_{q} E\big(u(q^{k})\big)
    &= \left( \frac{\partial u}{\partial q} \right)^{\!T}
       \nabla^{2}_{u} E\big(u(q^{k})\big)
       \left( \frac{\partial u}{\partial q} \right)
       = U^{T} \nabla^{2}_{u} E\big(u(q^{k})\big) U,
\end{align}
where $\nabla_{u} E$ and $\nabla^{2}_{u} E$ are the gradient and Hessian of the
energy with respect to the full displacement vector $u$.

After updating $q(t)$, the full state $u(t)$ is reconstructed and the simulation advances to the next time step.  
Importantly, the adaptive basis $U(t)$ is updated only once per time step while the inner Newton iterations for solving $q(t)$ proceed with this fixed basis.  
This separation keeps the per-step cost low and ensures stable convergence.  
The overall procedure for a single time step is summarized in Algorithm~\ref{alg:newton}.

\begin{algorithm}
\caption{Latent Space Dynamics via Newton's Method (one time step $t$)}
\label{alg:newton}
\begin{algorithmic}[1]
\STATE \textbf{Input:} Previous state $u(t-1)$, basis $U(t-1)\in\mathbb{R}^{3n\times r}$, energy $E(u)$
\STATE \textbf{Initialize:} $q^{0} \leftarrow 0$
\FOR{Newton iteration $k = 0,1,\ldots$}
    \STATE $u^{k} \leftarrow u(t-1) + U(t-1)\,q^{k}$
    \STATE $g^{k} \leftarrow U(t-1)^{\!T}\,\nabla_{u} E(u^{k})$
    \STATE $H^{k} \leftarrow U(t-1)^{\!T}\,\nabla^{2}_{u} E(u^{k})\,U(t-1)$
    \STATE Solve $H^{k}\,\Delta q^{k} = -\,g^{k}$
    \STATE Determine step size $\alpha^{k}\in(0,1]$
    \STATE $q^{k+1} \leftarrow q^{k} + \alpha^{k}\,\Delta q^{k}$
    \IF{convergence criterion satisfied}
        \STATE \textbf{break}
    \ENDIF
\ENDFOR
\STATE $q(t) \leftarrow q^{k+1}$,\quad $u(t) \leftarrow u(t-1) + U(t-1)\,q(t)$
\end{algorithmic}
\end{algorithm}

\subsection{Grassmann Interpolation for Large Deformation}
\begin{figure}[htbp]
  \centering
  \includegraphics[width=0.45\textwidth]{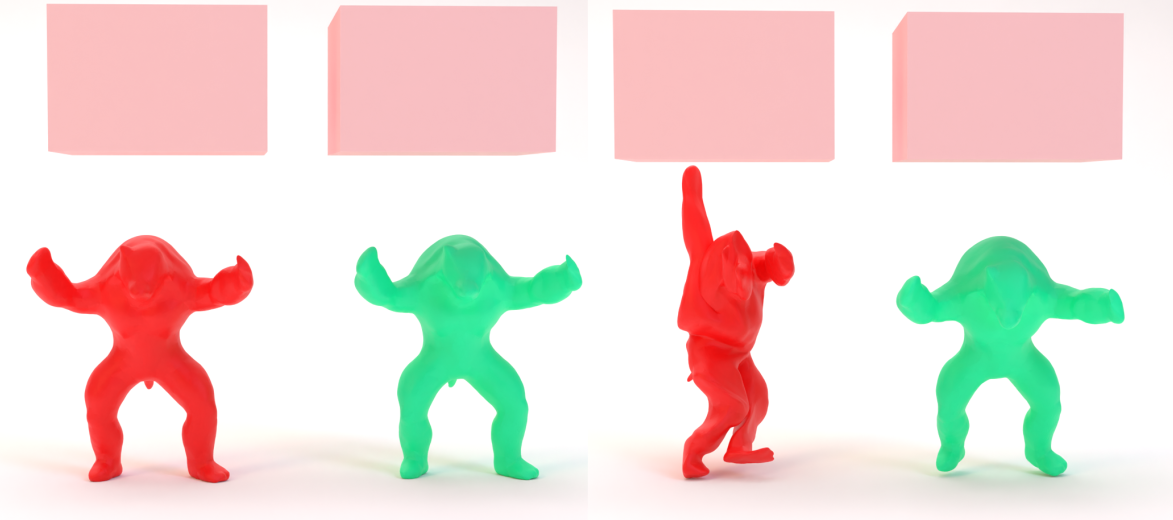}
  \caption{Comparison of adaptive linear model reduction under large    deformation. \textbf{Left}: Two identical objects before compression.
  \textbf{Right}: Results after compression—\color{red}without historical interpolation\color{black}\ and \color{ForestGreen}\ with historical interpolation\color{black}. The interpolation enables perfect recovery to the original state, whereas the standard one fails under large deformation.}
  \label{fig:large}
\end{figure}

While the adaptive linear model reduction performs well for small deformations, it encounters significant challenges under large deformations. As illustrated in Figure~\ref{fig:large}, the object may fail to recover its original state after experiencing substantial deformation. This issue arises because the current basis adaptation scheme does not account for the history of previous displacements, leading to a loss of important information about the deformation trajectory.

To address this limitation, we incorporate historical displacement information into the basis construction.  
We formulate the following least-squares problem:
\begin{equation}
    \min_{\Phi}\;\bigl\|\,\Phi P - D\,\bigr\|_F^2,
\end{equation}
where $\Phi \in \mathbb{R}^{3n \times r}$ is the basis matrix to be estimated,  
$P = [\,p(t\!-\!m\!+\!1)\;\big|\;\cdots\;\big|\;p(t)\,] \in \mathbb{R}^{r \times m}$ stacks the latent variables from the most recent $m$ time steps,  
and $D = [\,u(t\!-\!m\!+\!1)\;\big|\;\cdots\;\big|\;u(t)\,] \in \mathbb{R}^{3n \times m}$ collects the corresponding full displacements.  
Intuitively, this formulation seeks a basis that best captures the essential deformation patterns of the recent trajectory.  
When $\operatorname{rank}(P)=r$, the problem has a closed-form solution
\[
\Phi = D\,P^{\!T}\bigl(P P^{\!T}\bigr)^{-1}.
\]
To guard against rank deficiency, we regularize by adding a small diagonal term ($10^{-8}$ in our implementation) to $P P^{\!T}$.  
The computational cost of constructing $\Phi$ is dominated by the matrix multiplications and the inversion of the $r\times r$ system, yielding a per–time step complexity of $\mathcal{O}(n m r + r^3)$.

After obtaining the historical displacement basis $\Phi$, we blend it with the current adaptive basis $U$ using Grassmann interpolation to create the interpolated adaptive basis $U_\text{interp}$ as shown in Algorithm \ref{alg:grassmann}. Intuitively, Grassmann interpolation allows us to smoothly transition between two sets of basis vectors—effectively “morphing” one subspace into another in the most natural way, while always remaining within the space of valid bases. The interpolation is governed by a parameter $\lambda \in [0,1]$: when $\lambda=0$, the result is exactly the same span as historical basis $\Phi$; when $\lambda=1$, it is the same span as current basis $U$; and for intermediate values, the interpolated basis $U_{\text{interp}}$ combines information from both. This approach ensures that $U_{\text{interp}}$ not only adapts to the current deformation but also retains important features from the recent deformation history. By performing the interpolation on the Grassmann manifold—the mathematical space of all $r$-dimensional subspaces—this method preserves the geometric structure of the bases, resulting in more robust and physically meaningful model reduction, especially under large deformations.

The Grassmann interpolation procedure involves several matrix operations: computing the thin QR factorization, computing the singular value decomposition (SVD) of $U^T\Phi$, matrix multiplications, and elementwise trigonometric functions on the principal angles. The overall computational complexity for the interpolation step is $\mathcal{O}(nr^2)$, which is efficient in practice since $r \ll n$.

By incorporating historical information and employing Grassmann interpolation, our method effectively mitigates the limitations of purely adaptive bases and enables the system to recover from large deformations more reliably. The final algorithm is summarized in Algorithm~\ref{alg:final}.

\begin{algorithm}
\caption{Adaptive Linear Model Reduction with Grassmann Interpolation}
\label{alg:final}
\begin{algorithmic}[1]
\STATE \textbf{Input:} Initial displacement $u(0)\in\mathbb{R}^{3n}$
\FOR{each time step $t = 1,2,\ldots$}
    \STATE $p(t-1) \leftarrow e_{\theta_e}\big(u(t-1)\big)$
    \STATE $U(t-1) \leftarrow \dfrac{\partial g_{\theta_g}\big(p(t-1),\mathbf{x}\big)}{\partial p(t-1)}$
    \STATE Construct historical basis $\Phi$ using the most recent $m$ latent codes and displacements
    \STATE $U_{\mathrm{interp}}(t-1) \leftarrow \textsc{GrassmannInterpolation}\big(\Phi,\,U(t-1),\,\lambda\big)$ \quad (Algorithm~\ref{alg:grassmann})
    \STATE Compute $q(t)$ with Newton solver (Algorithm~\ref{alg:newton}) using basis $U_{\mathrm{interp}}(t-1)$
    \STATE $u(t) \leftarrow u(t-1) + U_{\mathrm{interp}}(t-1)\,q(t)$
\ENDFOR
\end{algorithmic}
\end{algorithm}

\begin{algorithm}
\caption{Grassmann Interpolation}
\label{alg:grassmann}
\begin{algorithmic}[1]
\STATE \textbf{Input:} Bases $\Phi, U \in \mathbb{R}^{3n \times r}$, interpolation parameter $\lambda \in [0,1]$ \hspace{-5mm}
\STATE \textbf{Output:} Interpolated basis $U_{\mathrm{interp}} \in \mathbb{R}^{3n \times r}$
\STATE Thin QR: $Q_\Phi R_\Phi \leftarrow \mathrm{qr}(\Phi)$,\; $Q_U R_U \leftarrow \mathrm{qr}(U)$
\STATE $M \leftarrow Q_\Phi^\top Q_U$
\STATE SVD: $M = V\,\cos\Theta\,W^\top$
\STATE $S \leftarrow \big(Q_U W - Q_\Phi V \cos\Theta\big)\,\big(\sin\Theta\big)^{-1}$
\STATE $G \leftarrow Q_\Phi V \cos(\lambda\Theta) + S \sin(\lambda\Theta)$
\STATE Thin QR: $U_{\mathrm{interp}} \leftarrow \mathrm{qr}(G)$
\RETURN $U_{\mathrm{interp}}$
\end{algorithmic}
\end{algorithm}

\section{Results}
We evaluate our method on a suite of benchmark deformable objects spanning a variety of shapes and material settings.
Training data are generated from long-duration dynamic simulations in which each object is repeatedly dropped into a Pachinko-style environment and tumbled in a dryer-like setting to induce a wide range of deformations, resulting in roughly $300{,}000$ displacement snapshots per object.
All models share a Young’s modulus of $10^{5}$ and a Poisson’s ratio of $0.45$.
For the CROM architecture, the encoder is a multilayer perceptron (MLP) with two hidden layers of $128$ units each, and the decoder has eight hidden layers of $128$ units; all hidden layers use ELU activations. 
Networks are trained with Adam and a scheduled learning rate that starts at $5$ and decreases to $2$, $1$, and $0.5$ over four phases of $100$ epochs each (total $400$ epochs), with batch size $64$ and GPU acceleration. 
The training loss is the mean squared error between reconstructed and ground-truth displacements.

After establishing the experimental setup, we present an ablation study of key parameters, a comparison of accuracy and visual quality against PCA-based ROM and the full-order model (FOM), and a runtime analysis.

\subsection{Ablation Study}

We investigate the influence of three key parameters: the number of basis vectors $r$, the history window size $m$, and the interpolation parameter $\lambda$. The parameter $r$ is standard in ROM, while $m$ and $\lambda$ are specific to our history-aware interpolation. For each configuration, we compute a relative online error by accumulating the per–time step $\ell_2$ displacement difference between each ROM (PCA and ours) and the FOM over a given simulation interval, and normalizing by the accumulated FOM displacement magnitude over the same interval. To prevent long-horizon drift from dominating the **quantitative** error measurements, the ROM state is realigned to the FOM state at the end of each time step before proceeding. This alignment is used only for numerical error evaluation and is not applied in the qualitative visual comparisons.

\subsubsection{Effect of Basis Size $r$}

\begin{figure}[t]
    \centering
    \begin{subfigure}[b]{0.48\textwidth}
        \centering
        \includegraphics[width=\linewidth]{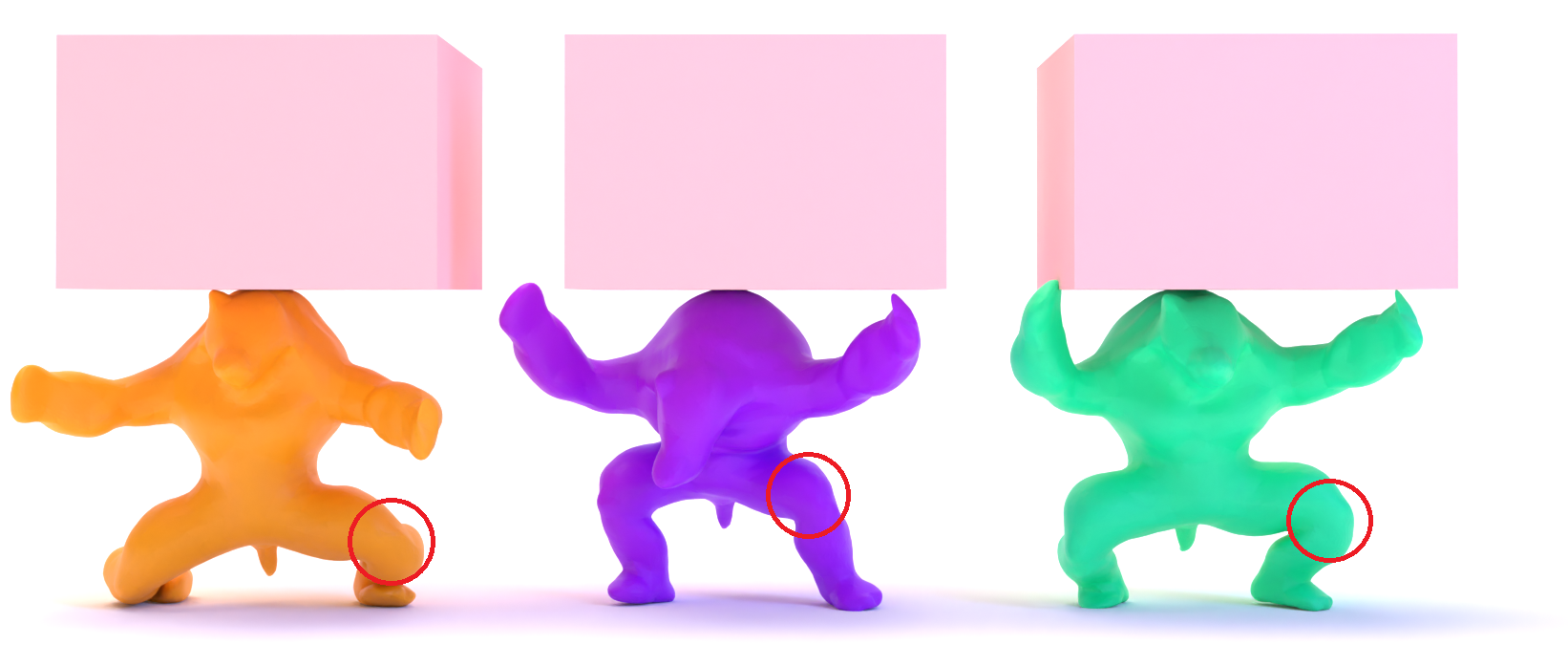}
        \caption{Armadillo Compression}
        \label{fig:arm_comp_img}
    \end{subfigure}
    \hfill
    \begin{subfigure}[b]{0.48\textwidth}
        \centering
        \includegraphics[width=\linewidth]{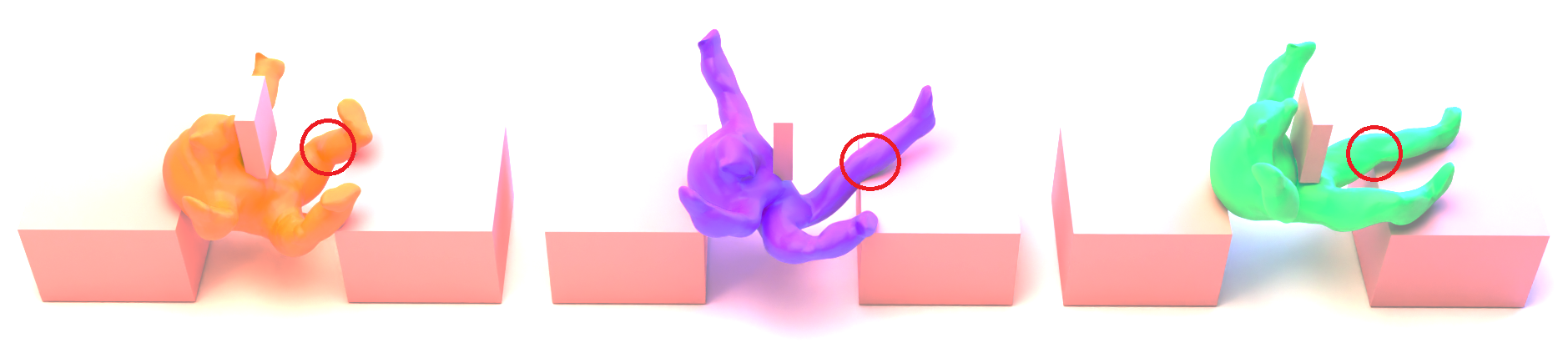}
        \caption{Armadillo Bending}
        \label{fig:arm_bend_img}
    \end{subfigure}
    \caption{Ablation study scenarios: Each subfigure shows three armadillos—\color{orange}FOM\color{black}, \color{violet}PCA\color{black}, and \color{ForestGreen}ours\color{black}. As highlighted by the red circle, \color{ForestGreen}ours\color{black}\ exhibits greater similarity to \color{orange}FOM\color{black}.
    }
    \label{fig:armadillo_compress_bending}
\end{figure}

\begin{figure}[t]
    \centering
    \begin{subfigure}[b]{0.48\textwidth}
        \centering
        \includegraphics[width=\linewidth]{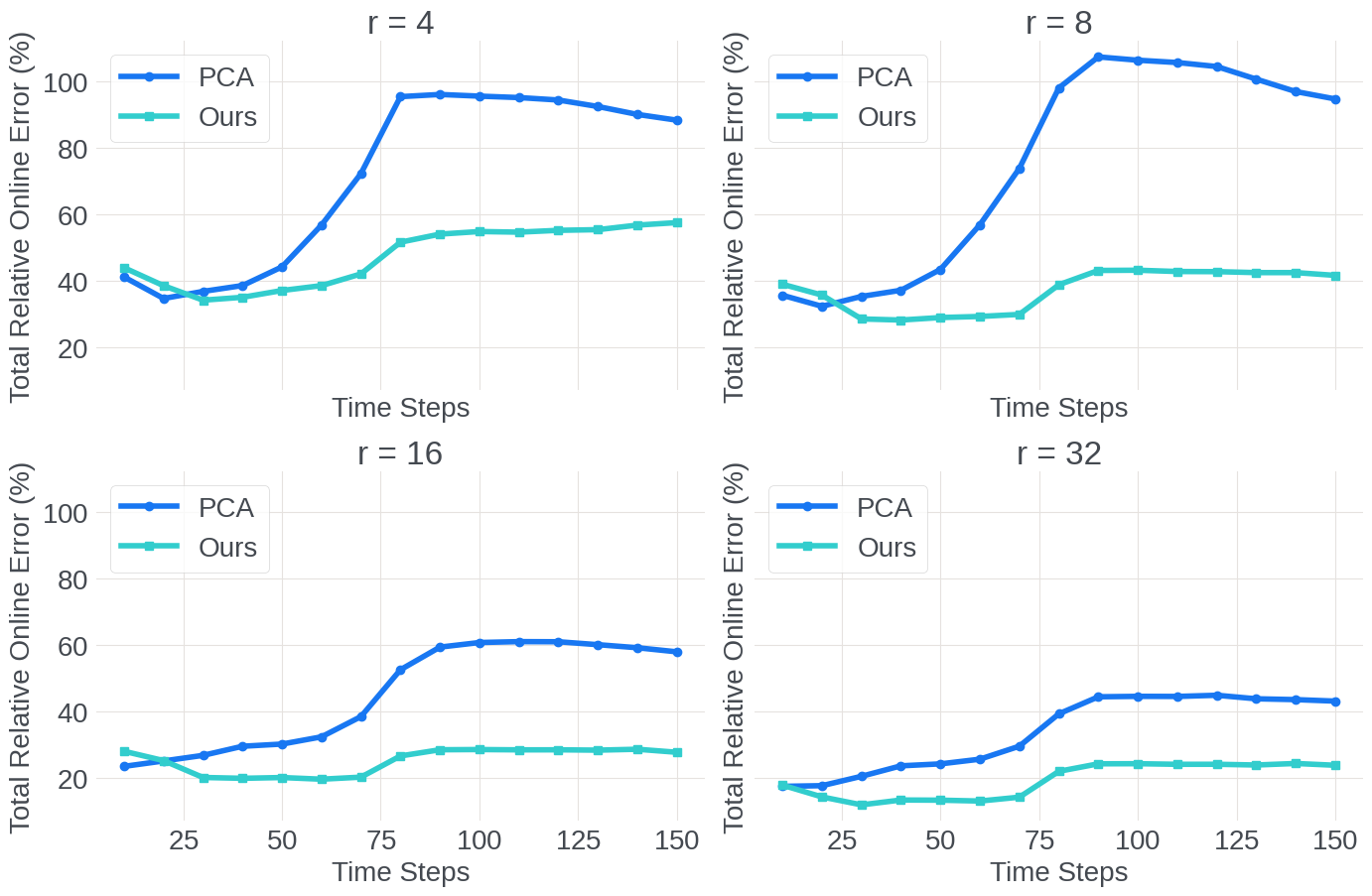}
        \caption{Armadillo Compression}
        \label{fig:arm_comp_err}
    \end{subfigure}
    \hfill
    \begin{subfigure}[b]{0.48\textwidth}
        \centering
        \includegraphics[width=\linewidth]{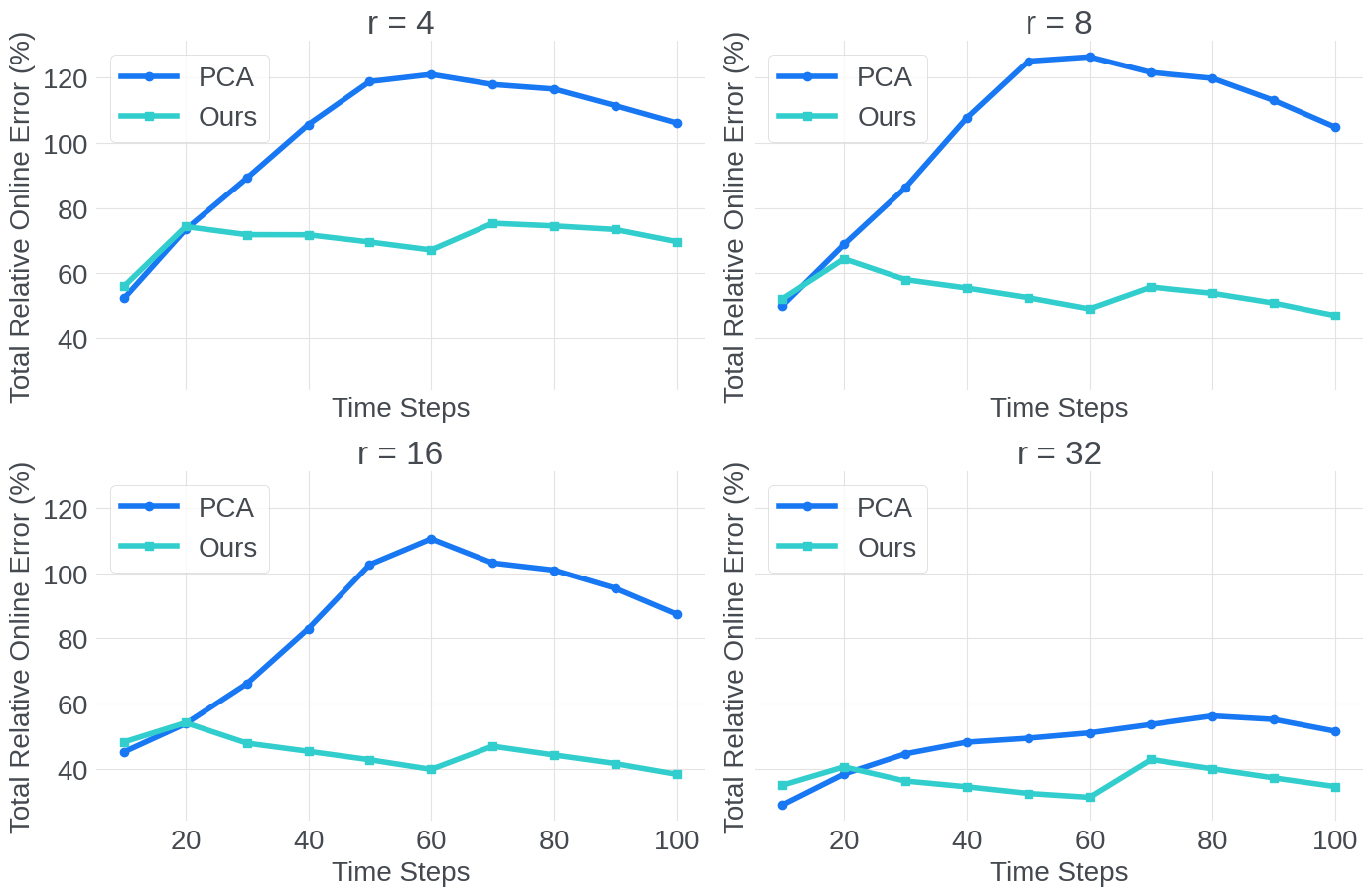}
        \caption{Armadillo Bending}
        \label{fig:arm_bend_err}
    \end{subfigure}
    \caption{Effect of basis size $r$ on relative online error for the two scenarios. Increasing $r$ improves accuracy for both methods and reduces the performance gap, but our approach remains consistently more accurate than PCA.}
    \label{fig:ablation_r}
\end{figure}

We assess sensitivity to $r$ on the armadillo compression and bending examples (Figure~\ref{fig:armadillo_compress_bending}). 
As shown in Figure~\ref{fig:ablation_r}, our method consistently achieves lower online error than PCA-based ROM for all tested $r \in \{4, 8, 16, 32\}$. 
While increasing $r$ improves both methods and narrows the gap, our approach remains more accurate across settings. 
We adopt $r=16$ as the default in subsequent experiments unless otherwise stated.

\subsubsection{Effect of History Window Size $m$}

We vary the history window $m$ to evaluate its effect on recovery from large deformations. Table~\ref{tab:ablation_m} reports total relative online error for Armadillo compression and bending. 
Very small windows (e.g., $m=1$) underperform due to insufficient history. 
Windows in the range $m\in[5,10]$ generally minimize error, though too-small $m$ can reduce visual smoothness. 
In practice we select $m=20$ as a good balance between accuracy and visual quality and use it by default thereafter.

\subsubsection{Effect of Interpolation Parameter $\lambda$}

The parameter $\lambda$ controls the blend between the historical basis and the current adaptive basis. 
Lower $\lambda$ values (more emphasis on history) tend to reduce error, but setting $\lambda$ too small (e.g., $\lambda=0$) can cause numerical issues by relying almost exclusively on historical information. 
Table~\ref{tab:ablation_lambda} summarizes results; we find $\lambda \in [0.1, 0.3]$ robust across examples and use $\lambda=0.2$ by default.

\begin{table}[t]
    \centering
    \begin{tabular}{c|cccccc}
        \toprule
        m & 1 & 5 & 10 & 20 & 30 & 50 \\
        \midrule
        \multicolumn{7}{c}{\textbf{Armadillo Compression}} \\
        \midrule
        $r=8$  & 78.90 & 31.84 & 32.18 & 41.77 & 45.79 & 58.57 \\
        $r=16$ & 77.10 & 29.26 & 26.33 & 27.80 & 29.77 & 37.36 \\
        \midrule
        \multicolumn{7}{c}{\textbf{Armadillo Bending}} \\
        \midrule
        $r=8$  & 88.25 & 38.25 & 40.50 & 47.03 & 56.12 & 75.55 \\
        $r=16$ & 74.60 & 37.56 & 36.32 & 38.31 & 43.97 & 62.98 \\
        \bottomrule
    \end{tabular}
    \caption{Total relative online error (\%) for different history window sizes $m$ on the armadillo compression and bending examples. Smaller windows lack sufficient history and yield higher errors, while moderate windows ($m\in[5,10]$) generally achieve the best accuracy. We adopt $m=20$ as a practical default to balance accuracy and visual smoothness.}
    \label{tab:ablation_m}
\end{table}

\begin{table}[t]
    \centering
    \begin{tabular}{c|cccccc}
        \toprule
        $\lambda$ & 0.0 & 0.1 & 0.2 & 0.3 & 0.4 & 0.5 \\
        \midrule
        \multicolumn{7}{c}{\textbf{Armadillo Compression}} \\
        \midrule
        $r=8$  & 36.12 & 36.72 & 41.77 & 45.97 & 49.21 & 55.08 \\
        $r=16$ & 40.08 & 24.93 & 27.80 & 30.24 & 34.95 & 38.39 \\
        \midrule
        \multicolumn{7}{c}{\textbf{Armadillo Bending}} \\
        \midrule
        $r=8$  & 44.58 & 44.48 & 47.03 & 51.37 & 56.80 & 63.94 \\
        $r=16$ & 36.37 & 37.63 & 38.31 & 42.01 & 45.52 & 50.47 \\
        \bottomrule
    \end{tabular}
    \caption{Total relative online error (\%) for different $\lambda$ values on the armadillo compression and bending examples. Smaller $\lambda$ values, which place more weight on the historical basis, generally reduced error, but setting $\lambda$ too close to zero can cause numerical instability. Values in the range $\lambda\in[0.1, 0.3]$ provide robust performance and we use $\lambda=0.2$ as the default.}
    \label{tab:ablation_lambda}
\end{table}

\subsection{Accuracy and Visual Quality}

\begin{figure}[htbp]
    \centering
    \begin{subfigure}[b]{0.48\textwidth}
        \centering
        \includegraphics[width=\linewidth]{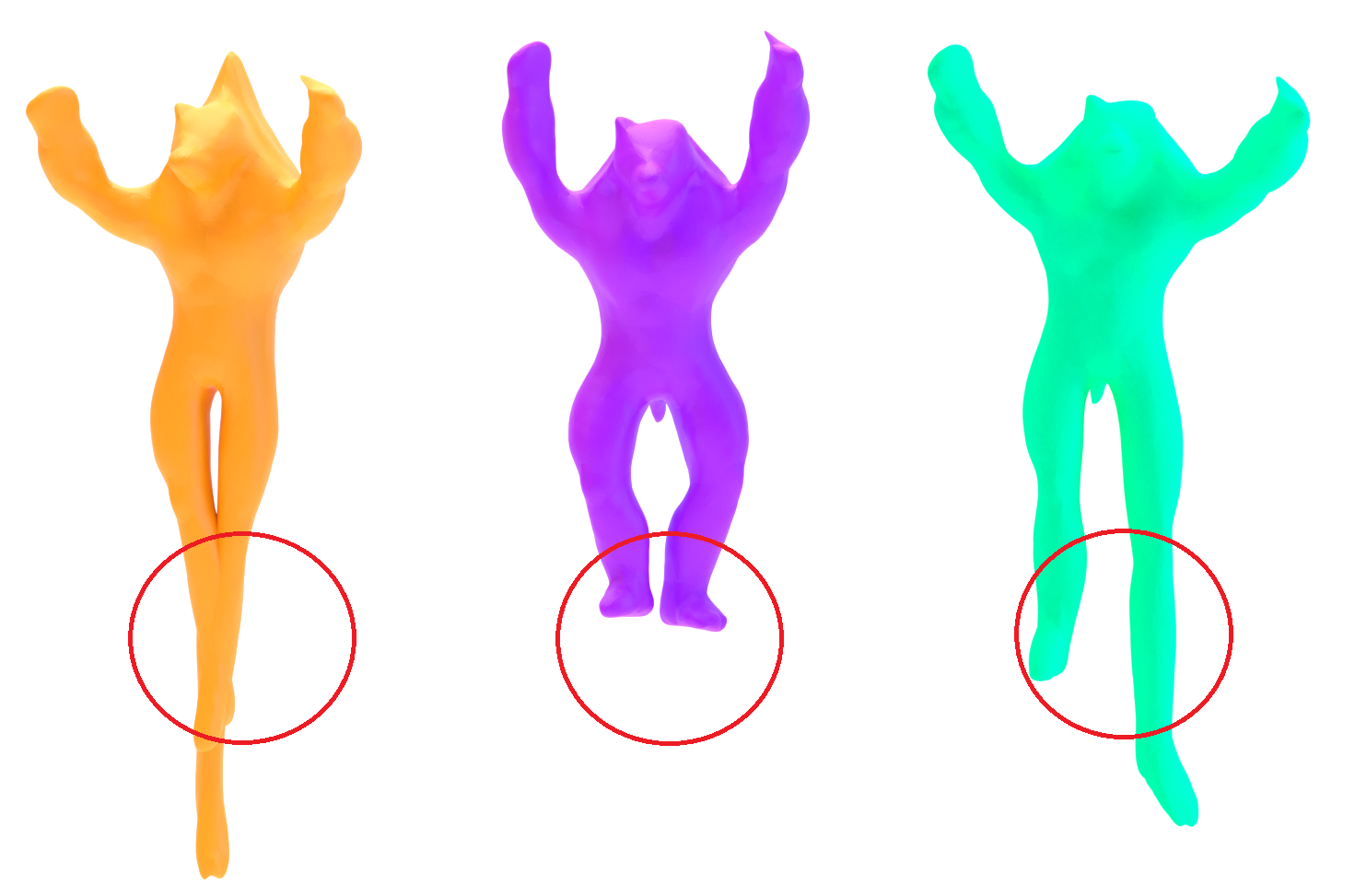}
        \caption{Armadillo Stretching}
        \label{fig:arm_stretch_img}
    \end{subfigure}
    \hfill
    \begin{subfigure}[b]{0.48\textwidth}
        \centering
        \includegraphics[width=\linewidth]{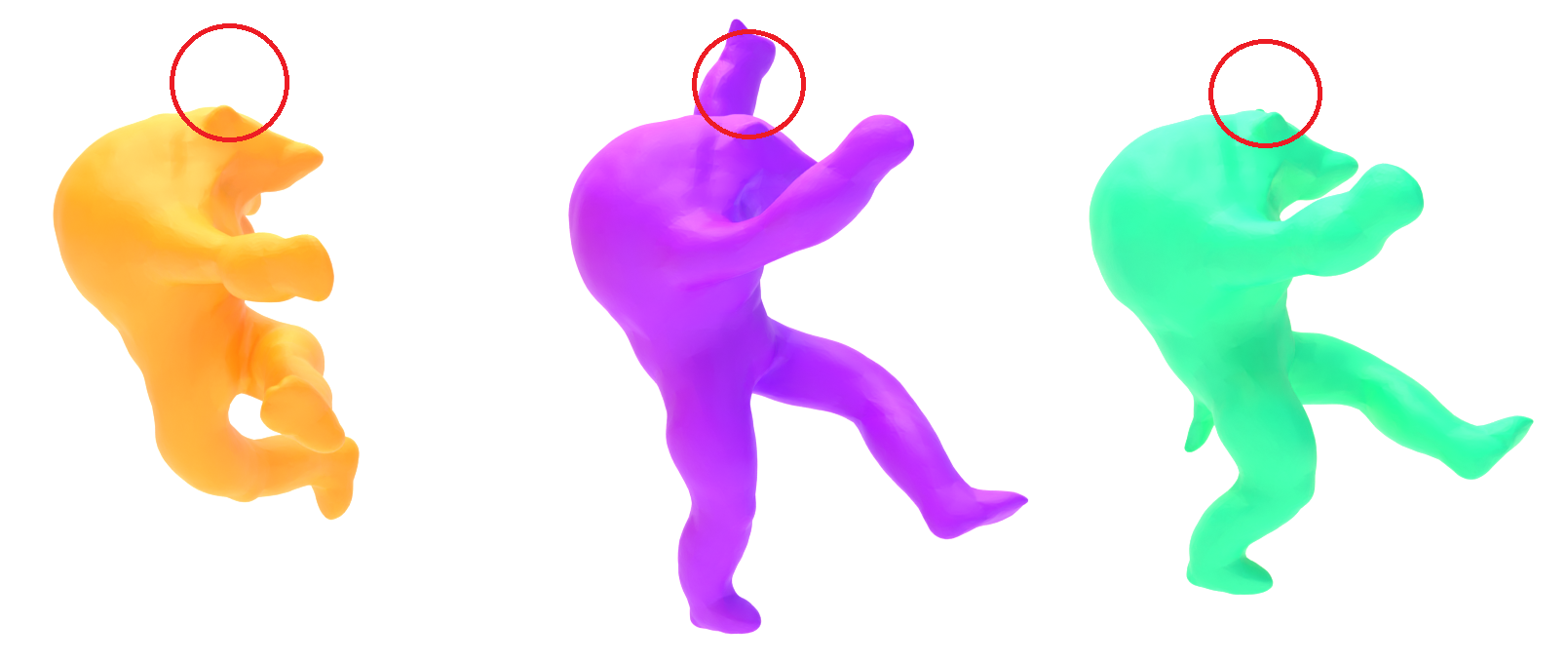}
        \caption{Armadillo Twisting}
        \label{fig:arm_twist_img}
    \end{subfigure}
    \hfill
    \begin{subfigure}[b]{0.48\textwidth}
        \centering
        \includegraphics[width=\linewidth]{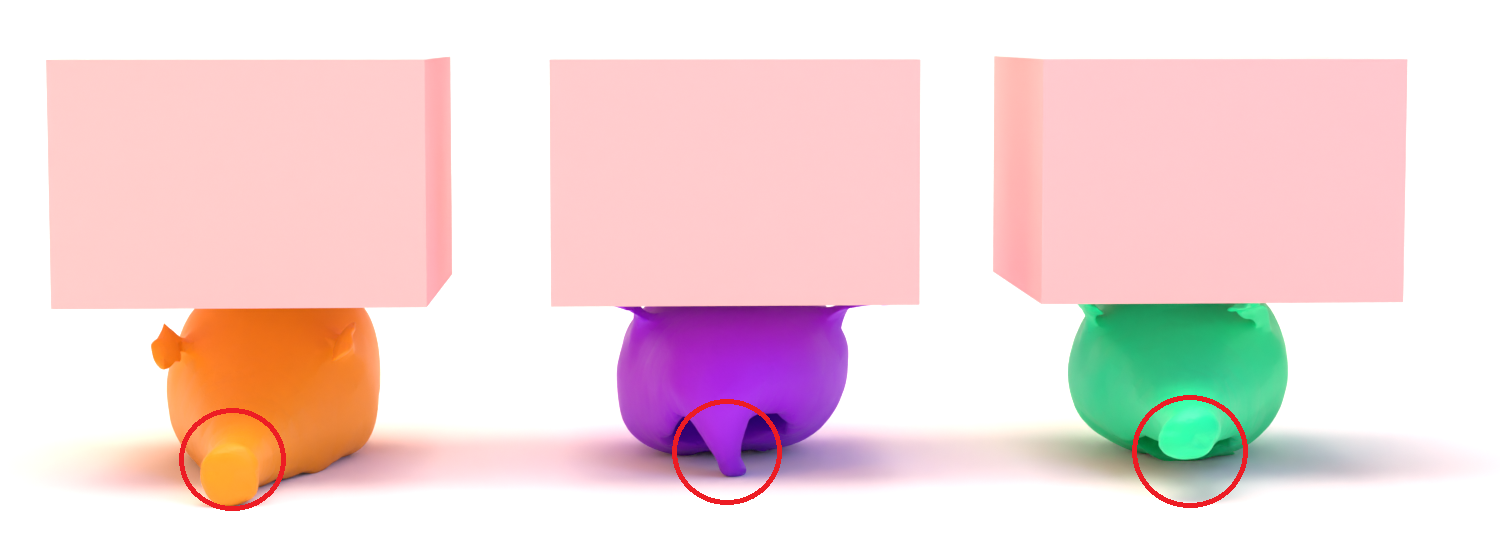}
        \caption{Pig Compression}
        \label{fig:pig_compress_img}
    \end{subfigure}
    \caption{Representative visual comparisons across scenarios: \color{orange}FOM\color{black}, \color{violet}PCA\color{black}, and \color{ForestGreen}ours\color{black}.}
    \label{fig:visual_comparison}
\end{figure}

We evaluate accuracy across a diverse set of objects and scenarios, comparing our approach to a PCA-based ROM baseline. 
The total relative online error is accumulated over the entire simulation, with per–step realignment to the FOM state to avoid divergence. 
Table~\ref{tab:accuracy_comparison} reports errors, showing that our method consistently outperforms PCA, particularly on challenging cases. 
Representative visual comparisons are given in Figure~\ref{fig:visual_comparison}.

\begin{table}[t]
    \centering
    \begin{tabular}{l l c c}
        \toprule
        Object & Scenario & PCA & Ours\\
        \midrule
        Armadillo & Compression & 58.02 & \textbf{27.80} \\
        Armadillo & Bending     & 87.44 & \textbf{38.31} \\
        Armadillo & Stretching  & 46.05 & \textbf{40.41} \\
        Armadillo & Twisting    & 31.09 & \textbf{21.59} \\
        Bunny     & Compression & 103.40 & \textbf{28.46} \\
        Bunny     & Bending     & 102.10 & \textbf{32.86} \\
        Dragon    & Stretching  & 37.68 & \textbf{17.9} \\
        Dragon    & Bending     & 75.28 & \textbf{40.67} \\
        Pig       & Compression & 69.53 & \textbf{29.62} \\
        Pig       & Stretching  & 95.13 & \textbf{86.64} \\
        \bottomrule
    \end{tabular}
    \caption{Total relative online error (\%) for multiple objects and scenarios. Our method reduces error by 40-70\% compared to PCA.}
    \label{tab:accuracy_comparison}
\end{table}

\subsection{Performance}

We assess computational efficiency by measuring the average runtime per simulation step across objects and scenarios, and compare against PCA-based ROM. 
All timings are recorded on a MacBook Pro with an M4 Pro processor, 48~GB RAM, and 4 threads. 
Results are summarized in Table~\ref{tab:performance_comparison}. 
Although our adaptive basis introduces additional computations (basis updates and Grassmann interpolation), the implementation maintains interactive performance, with per–step runtimes within a factor of roughly $1.3$–$1.5$ of PCA-based ROM.

\begin{table}[htbp]
    \centering
    \begin{tabular}{l l r c c}
        \toprule
        Object & Scenario & \#Tets & PCA & Ours \\
        \midrule
        Armadillo & Compression & 6440 & 6.42\,ms & 8.87\,ms \\
        Armadillo & Bending     & 6440 & 6.17\,ms & 8.44\,ms \\
        Armadillo & Stretching  & 6440 & 5.70\,ms & 8.01\,ms \\
        Armadillo & Twisting    & 6440 & 4.65\,ms & 7.08\,ms \\
        Bunny     & Compression & 6566 & 6.28\,ms & 8.64\,ms \\
        Bunny     & Compression & 6566 & 6.35\,ms & 8.50\,ms \\
        Dragon    & Stretching     & 3840 & 2.58\,ms & 5.66\,ms \\
        Dragon    & Bending     & 3840 & 3.41\,ms & 5.62\,ms \\
        Pig       & Compression & 10926 & 10.58\,ms & 14.10\,ms \\
        Pig       & Stretching  & 10926 & 9.18\,ms & 10.65\,ms \\
        \bottomrule
    \end{tabular}
    \caption{Average runtime per simulation step (milliseconds). Our method remains interactive while adding only a modest overhead of about 1.3x-1.5x compared to PCA.}
    \label{tab:performance_comparison}
\end{table}

\section{Conclusions \& Limitations}
In this paper, we presented an adaptive linear model reduction framework enhanced with history-aware Grassmann interpolation, enabling robust and efficient simulation of soft-body dynamics under large deformations. By blending the computational efficiency of classical linear subspaces with the flexibility of a time-updated basis, our method not only improves accuracy and robustness over standard PCA-based ROM, but also maintains the ability to recover the undeformed rest state even after extreme deformations.

Despite these advances, several limitations remain. First, like most learning-based methods, the effectiveness of our approach depends on the quality and diversity of the training data used to construct the CROM manifold and the adaptive basis, but this challenge is well recognized across the field and can be mitigated with broader data coverage. Second, while the historical basis captures recent deformation trajectories, it is not yet clear whether this is the most effective way to encode long-term deformation history, making it a natural target for future improvements. Finally, the adaptive basis is derived from the gradient of the CROM decoder rather than an explicitly optimized subspace, yet this approximation is extremely fast to compute and keeps the method practical even in challenging configurations.

\section{Acknowledgments}
We gratefully acknowledge the use of mesh models provided by the Stanford 3D Scanning Repository and Keenan Crane. These datasets were instrumental in the evaluation and visualization of our proposed method.

\bibliographystyle{eg-alpha-doi}
\bibliography{paper}
\end{document}